\documentclass{elsart}
\usepackage{graphicx}% Include figure files
\usepackage{dcolumn}% Align table columns on decimal point
\usepackage{bm}% bold math
\journal{Physics Letter B}
\begin{document}
\begin{frontmatter}
\title{Search for neutrinoless decays $\tau\to 3 \ell$} 
\begin{abstract}
 We have searched for neutrinoless $\tau$ lepton decays into three charged 
leptons using an 87.1 fb$^{-1}$ data sample collected with the 
Belle detector at the KEKB  $e^+e^-$ collider. Since the number of
signal candidate events is compatible with that expected from the
background, we set 90\% confidence level upper limits on the branching 
fractions in the range $(1.9-3.5) \times 10^{-7}$ for various
decay modes $\tau^- \to \ell^{-}\ell^{+}\ell^{-}$
where $\ell$ represents $e$ or $\mu$. 
\end{abstract}
\collab{Belle Collaboration}
  \author[Tohoku]{Y.~Yusa}, % Tohoku
  \author[Tohoku]{T.~Nagamine}, % Tohoku
  \author[Tohoku]{A.~Yamaguchi}, % Tohoku
  \author[KEK]{K.~Abe}, % KEK
  \author[TohokuGakuin]{K.~Abe}, % TohokuGakuin
% \author[TIT]{N.~Abe}, % TIT
  \author[KEK]{T.~Abe}, % KEK
  \author[KEK]{I.~Adachi}, % KEK
  \author[Tokyo]{H.~Aihara}, % Tokyo
  \author[Nagoya]{M.~Akatsu}, % Nagoya
% \author[Hiroshima]{M.~Asai}, % Hiroshima
  \author[Tsukuba]{Y.~Asano}, % Tsukuba
% \author[Toyama]{T.~Aso}, % Toyama
% ...
% Ljubljana
% \author[Lausanne]{D.~Z\"urcher}, % Lausanne
%\address[Aomori]{Aomori University, Aomori, Japan}
% ...
\address[KEK]{High Energy Accelerator Research Organization (KEK), Tsukuba, Japan}
% ...
\address[Nagoya]{Nagoya University, Nagoya, Japan}
% ...
\address[TohokuGakuin]{Tohoku Gakuin University, Tagajo, Japan}
\address[Tohoku]{Tohoku University, Sendai, Japan}
\address[Tokyo]{Department of Physics, University of Tokyo, Tokyo, Japan}
% ...
\address[Tsukuba]{University of Tsukuba, Tsukuba, Japan}
% ...
%\thanks[Fermilab]{on leave from Fermi National Accelerator Laboratory, Batavia, IL, USA}
\thanks[NovaGorica]{on leave from Nova Gorica Polytechnic, Nova Gorica, Slovenia}
\end{frontmatter}

\section{Introduction}
 In the Standard Model (SM), lepton-flavor-violating (LFV) decays of 
charged leptons are forbidden or highly suppressed even if 
 neutrino mixing is taken into account~\cite{SMLFV}. LFV is expected 
in many extensions of the SM such as SUSY models with 
Higgs mediation~\cite{HIGGSM}, right-handed 
neutrinos~\cite{SUSYNR1,SUSYNR2}, multi-Higgs bosons~\cite{MHIGGS}, 
extra $Z'$ gauge bosons~\cite{ZPRIM} and $R$-parity violating 
interactions~\cite{RVIO}. 
Some  of those models predict LFV decays of charged $\tau$ leptons
enhanced to a level accessible at present $B$-factories. 
Observation of LFV would provide evidence for new physics 
beyond the SM.

In this paper, we report on a search for six LFV $\tau$ decay modes:
$\tau^{-} \to  e^{-} e^{+} e^{-}$, 
$\tau^{-} \to  e^{-} \mu^{+} \mu^{-}$,   
$\tau^{-} \to  e^{+} \mu^{-} \mu^{-}$,
$\tau^{-} \to  \mu^{-} e^{+} e^{-}$, 
$\tau^{-} \to  \mu^{+} e^{-} e^{-}$ and 
$\tau^{-} \to  \mu^{-} \mu^{+} \mu^{-}$.
Charge conjugate decay modes are implied throughout the paper.
Upper limits on the branching fractions for these decays 
at the level $(1-2) \times 10^{-6}$ at 90\% confidence level
have been set by the CLEO Collaboration using a data sample of 
4.79 fb$^{-1}$~\cite{CLEO3L}. Recently these results were
improved on by the BaBar experiment which reported upper limits in the range 
$(1.1-3.3) \times 10^{-7}$ from a 
91.5 fb$^{-1}$ data sample~\cite{BABAR3L}. We present here 
a new search based on a data sample of 87.1 fb$^{-1}$ corresponding to
(79.3 $\pm$ 1.1) million $\tau$-pairs collected with the Belle
detector~\cite{BELLE} 
at the KEKB asymmetric energy $e^{+}e^{-}$ collider~\cite{KEKB} 
operating at a center-of-mass energy $\sqrt{s}\simeq$ 10.6 GeV. 

\section{Event Selection}
 The Belle detector is a general purpose detector with excellent 
capabilities for precise vertex determination and particle 
identification. Tracking of charged particles is performed by a 
three-layer double-sided silicon vertex detector (SVD) and a 
fifty-layer cylindrical drift chamber (CDC) located in a 1.5 T 
magnetic field. Charged hadrons are identified by means of $dE/dx$ 
from the CDC, signal pulse-heights from aerogel $\check{\rm C}$erenkov 
counters (ACC), and timing information from time-of-flight 
scintillation counters (TOF). Energies of photons are 
measured using a CsI (Tl) electromagnetic calorimeter (ECL). Muons are 
detected by fourteen layers of resistive plate counters interleaved 
with iron plates (KLM). 

 We search for $\tau^{+}\tau^{-}$ events in which one $\tau$ decays into three 
charged leptons (3-prong) and the other $\tau$ decays into one charged 
and any number of neutrals (1-prong), which has a branching fraction of 
(85.35$\pm$0.07)\%~\cite{PDG}. We start by
requiring signal candidate events to have four charged tracks with 
zero net charge and any number of photons.
 Each charged track must have transverse momentum 
$p_{\rm t} > 0.1$ GeV/$c$ and be within the polar angle range 
25$^\circ < \theta < 140^\circ$. For each charged track,
the distance of closest approach to 
the interaction point (IP) is required to be within $\pm 1$ cm transversely 
and $\pm 3$ cm along the beam. Photon 
candidates are selected from neutral ECL clusters with an energy 
$E_{\rm cluster} >$ 0.1 GeV, where neutral ECL clusters must be separated 
by at least 30 cm from a projection point of any charged track in the 
ECL. 
 The tracks and photons in an event are divided into two hemispheres 
in the $e^+e^-$ center-of-mass system (CMS) by the plane 
perpendicular to the thrust axis calculated from the momenta of 
all charged tracks and photons in the event. Signal candidates have a 
1-prong vs 3-prong topology, i.e. three charged tracks are required in one
hemisphere, one charged track in the other. We define the former 
 hemisphere as the signal side and the latter as the tag side. The 
number of photon candidates on the signal side, $n_{\gamma}$, 
should be less than or equal to two, to allow for photons from initial 
state radiation or photon radiation in the detector by electron tracks. 

 Electrons are identified by means of an electron likelihood 
function ($\mathcal{L}_{e}$), that includes the information on 
the $dE/dx$ measurement in the CDC and the ratio of the cluster energy 
in the ECL to the track momentum measured in the CDC~\cite{eid-NIM}. 
For electrons, we require $\mathcal{L}_{e} > 0.1$ and the laboratory 
momentum to be greater than 0.3 GeV/$c$.
 In order to correct for the energy loss 
from bremsstrahlung in the detector material, the momentum 
of an electron candidate is recalculated by adding the momentum of 
radiated photon clusters when an ECL cluster with energy less 
than 1.0 GeV is detected 
within a cone angle of 10$^\circ$ around the flight direction 
of the electron candidate track.
The muon likelihood function ($\mathcal{L}_{\mu}$) is evaluated from 
two variables -- the difference between the range calculated from 
the momentum of a particle and the range measured in the KLM as well as 
the $\chi^2$ value of the KLM hits with respect to the extrapolated 
track~\cite{muid-NIM}. For muons, $\mathcal{L}_{\mu}$ 
is required to be larger than 0.1 and its momentum should be larger 
than 0.6 GeV/$c$. 
The requirement of lepton identification for all three charged tracks 
on the signal side leads to significant background reduction by a 
factor in the range from $10^2$ to $10^4$ for each decay mode 
with a 40\% loss of signal. 

For the Monte Carlo (MC) simulation of the signal, one $\tau$ lepton is
assumed to decay into the signal LFV modes with a uniform phase
space distribution in the $\tau$ rest frame and the other $\tau$ 
decays generically using the KORALB/TAUOLA program \cite{KORALB}.

One important source of background remaining after requiring the
event topology and lepton identification on the signal side
is due to radiative Bhabha events with a converted photon. 
This background is efficiently reduced by requiring that the invariant mass 
of any two oppositely charged particles be greater than 0.2
GeV$/c^{2}$, assuming the electron mass for each particle. The
remaining background after removing photon conversions 
comes from two-photon processes, $\tau$-pair 
events with generic decays into three charged hadrons, 
$e^+e^- \to q\bar{q}$ continuum and 
$B\bar{B}$ events, where hadrons are misidentified as 
leptons on the signal side. 

\begin{figure}[!ht]
\begin{center}
\includegraphics[scale=0.45,clip]{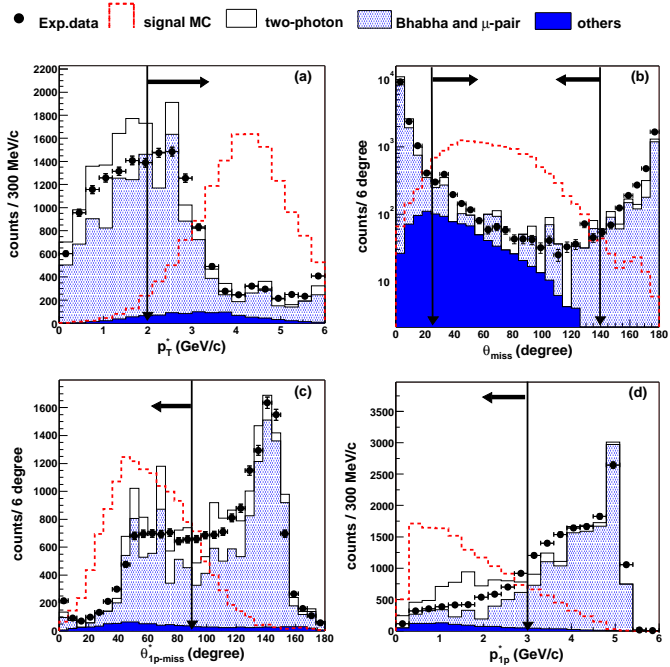}
\caption{ Kinematical distributions used in the event selection after 
the event topology and lepton identification requirements:
(a) the total transverse momentum $p_{\rm T}^*$,
(b) the polar angle of the missing momentum $\theta_{\rm miss}$,
(c) the opening angle between the momentum of a 1-prong track and 
the missing momentum $\theta^{*}_{\rm 1p-miss}$,
(d) the momentum of a 1-prong track $p^{*}_{\rm 1p}$.
The distributions for experimental data, signal MC and background MC 
are indicated by closed circles, the dashed and solid histograms, 
respectively. The areas of the signal MC are 
normalized assuming a branching fraction of $1.3 \times 10^{-3}$ 
while the background MC is normalized to the data luminosity. 
Background MC from two-photon, Bhabha and $\mu$-pair, and 
other processes (generic $\tau\tau$, $e^+e^- \to q\bar{q}$ continuum and $B\bar{B}$) are 
shown by the open, shaded and dark histograms, respectively.
The distributions are shown for the $\tau^- \to e^-e^+e^-$ mode.
Although the size of each background source depends on the decay mode, 
the shape of each background is common to all.
The shape of signal distributions is also similar for the six decay modes.
}
\label{Select}
\end{center}
\end{figure}

For signal $\tau$-pair events there is a missing momentum due to 
neutrino emission from the $\tau$ on the tag side. 
Figure~\ref{Select} (a) shows the total transverse momentum
$p^*_{\rm T}$, the transverse component of the sum of momentum vectors for 
the four charged tracks in the CMS. To suppress events from 
two-photon processes, $p^*_{\rm T}$ is required to be larger 
than 2.0 GeV/$c$. We calculate the missing momentum by 
subtracting the momentum of 
all charged tracks and photons from the beam momentum.
In order to suppress radiative Bhabha and two-photon events, 
the polar angle of the missing momentum in 
the laboratory frame $\theta_{\rm miss}$, must be between $25^{\circ}$ and 
$140^{\circ}$, as shown in Fig.~\ref{Select} (b). 
The missing momentum due to neutrinos from the $\tau$ on the tag side 
tends to lie in the same hemisphere as the 1-prong track for signal events. 
The opening angle between the 1-prong track and 
missing momentum in the CMS, $\theta^{*}_{\rm 1p-miss}$, is required to be 
less than 90$^\circ$, as shown in Fig.~\ref{Select} (c).  
Finally, the momentum of the 1-prong tag side track, 
$p^{*}_{\rm 1p}$, must be less than 3 GeV$/c$. Since the 
tag side $\tau$ decays with neutrino(s) or $\pi^0$ emission, the 
momentum of the 1-prong track is much smaller than the $\tau$ 
momentum. This requirement suppresses most of the Bhabha and $\mu$-pair 
backgrounds, as shown in Fig.~\ref{Select} (d).  
After all these requirements, the background is reduced by a 
factor of order $10^4$ with a 10\% efficiency for the signal.

\begin{figure}[ht!]
\begin{center}
\includegraphics[scale=0.5,clip]{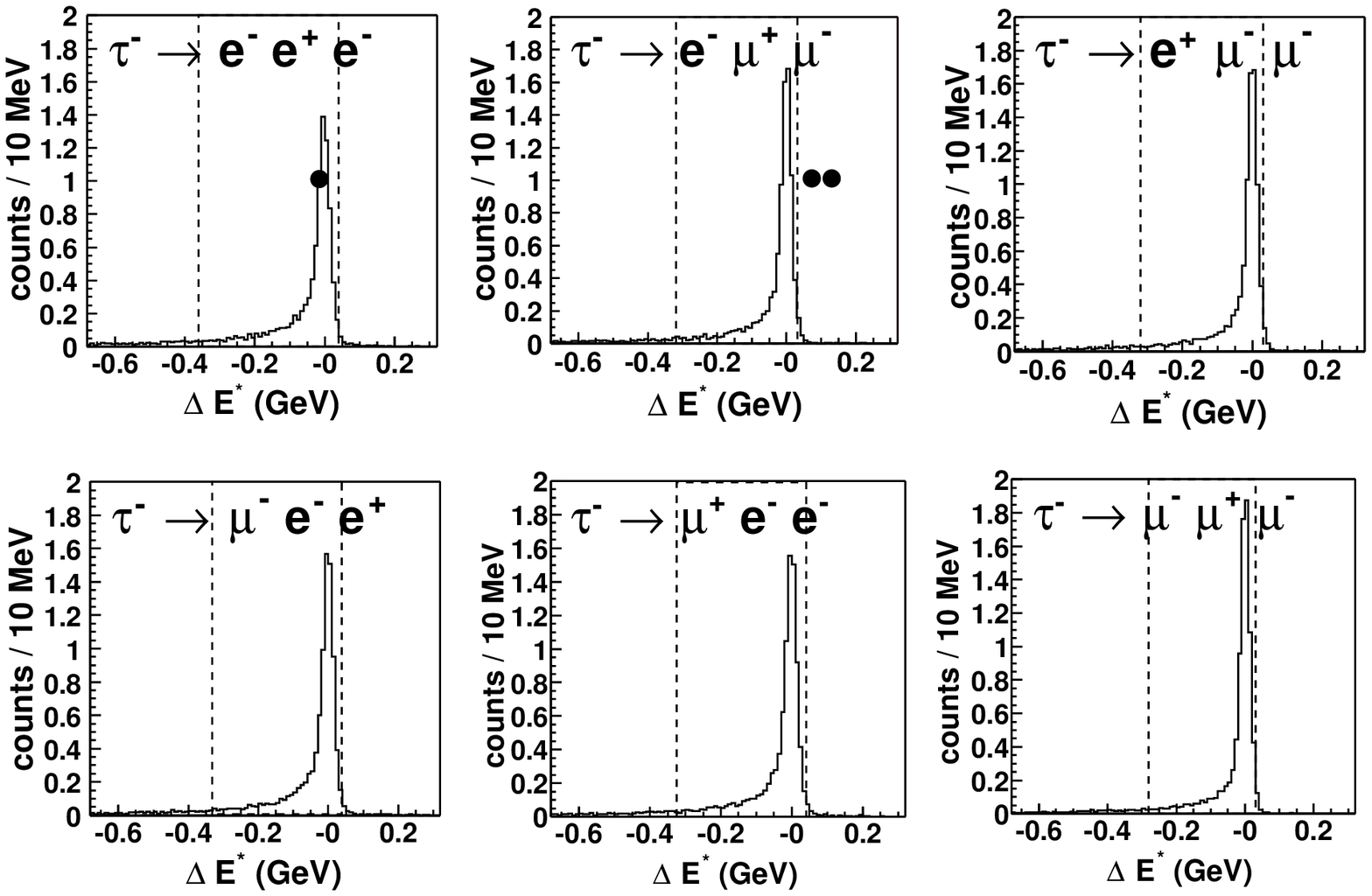}
\caption{$\Delta E^*$ distributions of the events in the $\Delta M$ 
signal region. Open histograms are the signal MC sample 
assuming the branching fraction of $1.3 \times 10^{-6}$, 
while experimental data is plotted by closed circles. The dashed lines 
indicate the $\Delta E^*$ signal region.}
\label{Ecmmc}
\includegraphics[scale=0.5,clip]{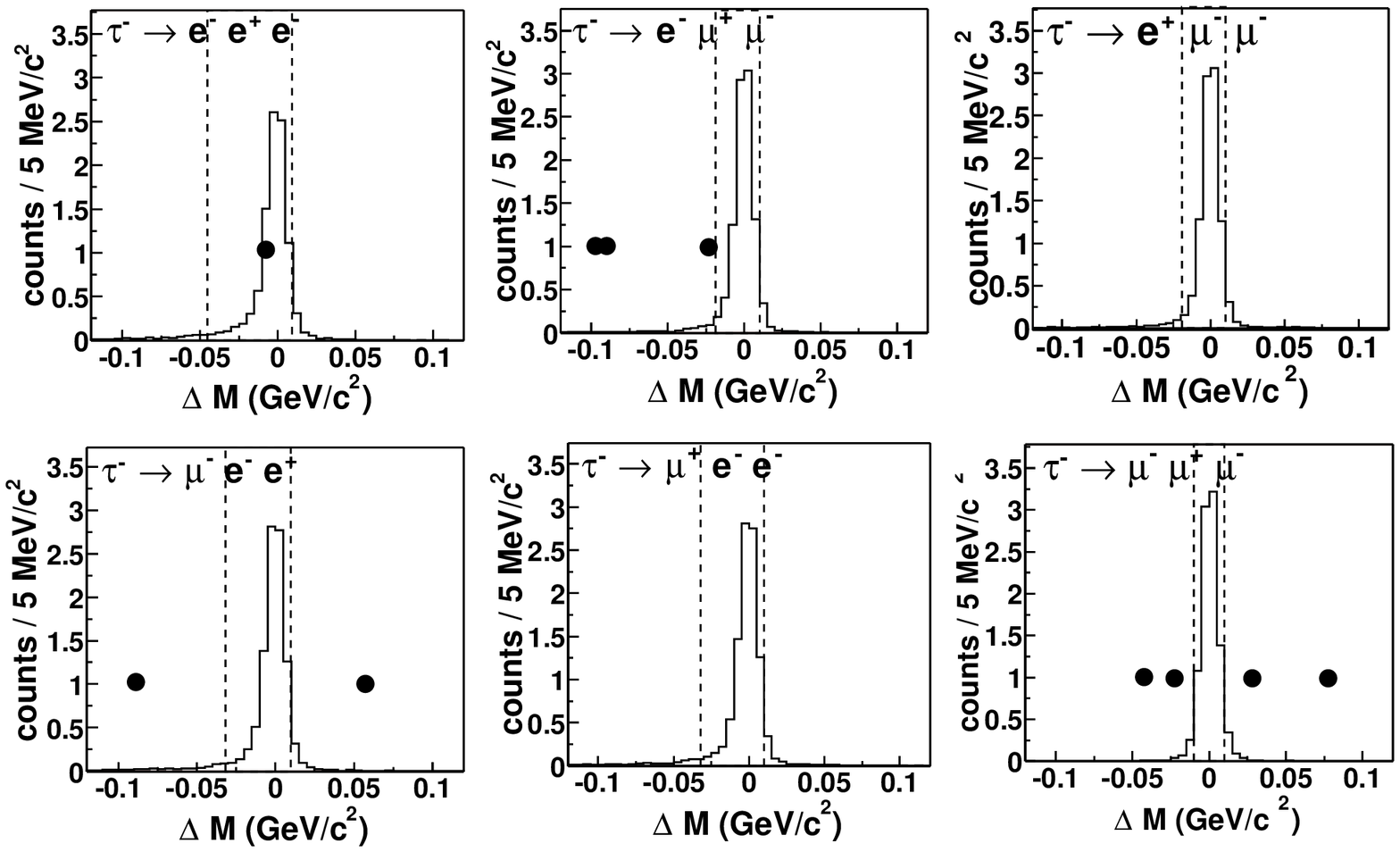}
\caption{$\Delta M$ distributions of the events in the $\Delta E^*$ 
signal region. Open histograms are the signal MC sample 
assuming the branching fraction of $1.3 \times 10^{-6}$, while
experimental data is plotted by closed circles. The dashed lines 
indicate the $\Delta M$ signal region.}
\label{Mmc}
\end{center}
\end{figure}

 From the total CMS energy, $E^*_{3\ell}$, and invariant mass, $M_{3\ell}$,
of the three signal side leptons, we compute the quantities: 
$\Delta E^* \equiv E^*_{3\ell} - E^{*}_{\rm beam}$ and
$\Delta M \equiv M_{3\ell} - M_{\tau}$, 
where $E^{*}_{\rm beam}$ is the CMS beam energy and $M_{\tau}$ is 
the $\tau$ mass. In the $\Delta E^{*}$-$\Delta M$ plane, the 
neutrinoless $\tau$ decay events are expected to be distributed close 
to the origin. The $\Delta E^*$ and $\Delta M$ distributions for each
decay mode are shown in Figs.~\ref{Ecmmc} and \ref{Mmc}, respectively
together with expectations based on signal MC.
It is seen that each peak has a tail on the lower side 
that is due to initial state radiation or bremsstrahlung of a 
charged particle interacting with the detector material. Since 
electrons radiate more than muons, the shape of the 
peak depends on decay mode. 

The signal region for each decay mode is given in Table~\ref{tab:signalerea}, 
and is illustrated as the region between two dashed lines in Figs.~\ref{Ecmmc} 
and \ref{Mmc}.
Each signal region is defined to contain 90\% of the signal MC events 
plotted in the figures. 
Figure~\ref{EvsMdata} shows the $\Delta E^*$ vs $\Delta M$ plot for
the data. In the fourth and fifth columns of Table~\ref{tab:signalerea}, 
we compare the number of events found for the data 
and the normalized MC background in the plotted region of 
$-0.68~ {\rm GeV} < \Delta E^* < 0.32 ~{\rm GeV}$ 
and $-0.12~ {\rm GeV}/c^{2}< \Delta M  < 0.12~{\rm GeV}/c^2$ 
shown in Fig.~\ref{EvsMdata}.

\section{Results}
Efficiencies for $\tau \to 3\ell$ decays with a uniform
phase space distribution vary from 9.2\% to 9.5\%
and are listed in the second column of Table~\ref{tab:result}.
The actual decay angle distribution, however, will depend on the LFV
interaction and include spin correlations between
the tag side and signal side $\tau$ \cite{OKADA}.
In order to evaluate the maximum possible effect of such correlations, 
we examine $V-A$ and $V+A$ interactions using the 
formulae given in \cite{OKADA}. 
The relative differences in the efficiencies ($\Delta\epsilon/\epsilon$) 
from a uniform distribution are found to be 
$-3.8\%, -8.7\%, -1.1\%, +0.8\%, -12.6\%$ and $-5.6\%$ for 
$\tau^- \to e^- e^+ e^-$, $\tau^- \to e^- \mu^+ \mu^-$, 
$\tau^- \to e^+ \mu^- \mu^-$, $\tau^- \to \mu^- e^- e^+$, 
$\tau^- \to \mu^+ e^- e^-$ and $\tau^- \to \mu^-\mu^+\mu^-$ 
decay mode, respectively. 
Requirements on the number of CDC tracks and 
the energy of ECL clusters that are 
used to detect $\tau$-pair events also constitute part of the trigger 
logic. The impact of the trigger on the efficiency is investigated by 
applying a trigger simulation to the signal MC.
The changes in the detection efficiencies are found to be only 
a few percent, because the selection criteria applied in this 
analysis are much more restrictive than the trigger conditions.

\begin{figure}
\begin{center}
\includegraphics[scale=0.7,clip]{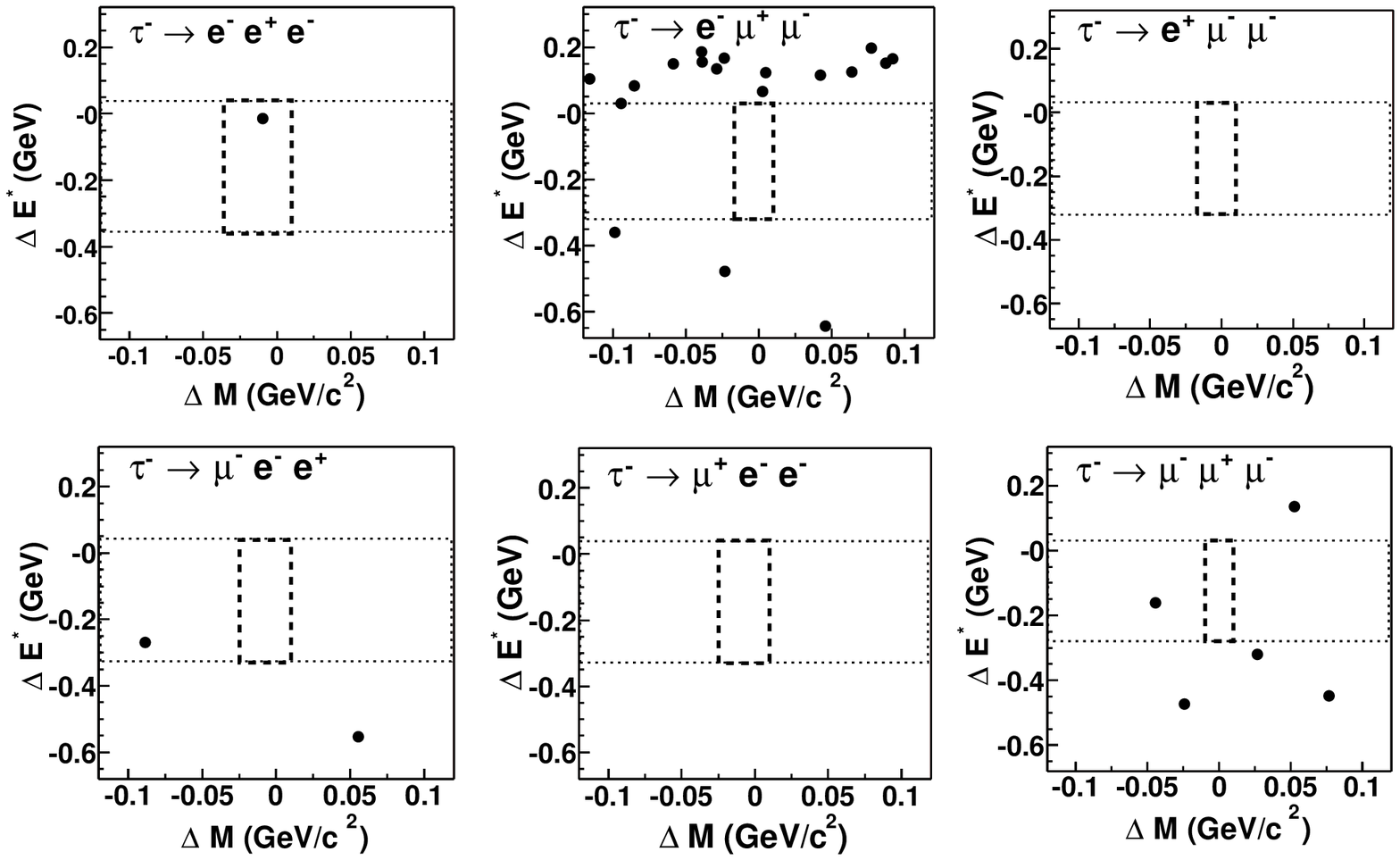}
\caption{
$\Delta E^*$ vs $\Delta M$ plots for the experimental data. The charge
conjugate decay mode is also included. The dashed and dotted boxes 
indicate the signal and $\Delta M$ side-band regions, respectively.
}
\label{EvsMdata}
\end{center}
\end{figure}

\begin{table}
\caption{ Definition of the signal regions for each decay mode and the 
number of events in Fig.~\ref{EvsMdata} for the data and the
normalized background MC.}
 \begin{tabular}{c c c c l }
  \hline
  \hline
Mode &\multicolumn{2}{c}{Signal region} & \multicolumn{2}{c}{Number of
events} \\
& & &\multicolumn{2}{c}{in the Fig.~\ref{EvsMdata} area}\\
  & $\Delta E^{*}$ (GeV) &  $\Delta M$  (GeV/$c^{2}$) & Data & MC\\
\hline
$\tau^-\to e^- e^+ e^-$     & $-0.36 < \Delta E^{*} < 0.04$  &
$-0.032 <\Delta M < 0.010$  & 1   & $0 ~^{+17}_{-0}$ \\
  \hline
$\tau^-\to e^-\mu^+\mu^-$   & $-0.32 < \Delta E^{*} < 0.03$  &
$-0.017 <\Delta M < 0.010$  & 18  & $5 ~^{+17}_{-3}$\\  
  \hline
$\tau^-\to e^+\mu^-\mu^-$   & $-0.32 < \Delta E^{*} < 0.03$  & 
$-0.017 <\Delta M < 0.010$  & 0   & $2 ~^{+3}_{-2}$ \\
  \hline
$\tau^-\to \mu^-e^-e^+$     & $-0.33 < \Delta E^{*} < 0.04$  & 
$-0.025 <\Delta M < 0.010$  & 2   & $0 ~^{+3}_{-0}$\\         
   \hline
$\tau^-\to \mu^+e^-e^-$     & $-0.33 < \Delta E^{*} < 0.04$  & 
$-0.025 <\Delta M < 0.010$  & 0   & $0 ~^{+3}_{-0}$\\      
  \hline
$\tau^-\to \mu^-\mu^+\mu^-$ & $-0.28 < \Delta E^{*} < 0.03$  & 
$-0.010 <\Delta M < 0.010$  & 5   & $8\pm 4$ \\  
  \hline
  \hline
 \end{tabular}\\
\\
\label{tab:signalerea}
\end{table}

As shown in Fig.~\ref{EvsMdata}, the background level in and 
around the signal region is very low.
From the background MC, we find that the remaining events are 
due to the low-multiplicity $e^+e^- \to q\bar{q}$ continuum or 
$B\bar{B}$ events where final-state hadrons are misidentified as 
leptons as well as a few events from generic $\tau$-pair decay. 
As seen from Table~\ref{tab:signalerea}, the numbers of events found 
in the Fig.~\ref{EvsMdata} area are consistent with the 
numbers expected from the normalized background MC. 

 To evaluate the background $b$ in the signal region, we assume a 
uniform background distribution along the $\Delta M$ axis in 
Fig.~\ref{EvsMdata}. With looser selection criteria, we find 
the $\Delta M$ distribution is uniform. 
We estimate the number of background events in the signal region 
from the number of events observed in the dotted box, the 
$\Delta M$ side-band regions in Fig.~\ref{EvsMdata}. 
The systematic uncertainty for this method is estimated by 
comparing the observed and estimated number of events in the 
region outside of the dotted box in Fig.~\ref{EvsMdata}. The fourth 
column of Table~\ref{tab:result} shows the number of estimated 
background events and its error, including both systematic and 
statistical uncertainties. 

In the signal regions for the six decay modes considered, one candidate is
observed for the $\tau^- \to e^-e^+e^-$ mode while no candidates
are found for the other modes. 
The numbers of events in the signal regions are consistent with
the background expectations.

\begin{table}
\caption{Summary of detection efficiency, number of observed events, 
background expectation,
$s_{0}$ and 90\% C.L. upper limits on the branching fractions.}
 \begin{tabular}{lccccc}
  \hline
  \hline
Decay     & Detection  & Number of  & Expected &  &  Upper limit \\
mode      & efficiency  & events observed   & background &$s_{0}$  & of $\mathcal{B}$\\
          &  ($\epsilon$),\ \%           &       &$(b)$  &  & $\times 10^{-7}$ \\
  \hline
$\tau^- \to e^- e^+ e^-$       & $9.2 \pm 0.6$ & 1 & $< 0.2$     & 4.36 & 3.5\\
  \hline
$\tau^- \to e^- \mu^+ \mu^-$   & $9.2 \pm 1.4$ & 0 & 0.1$\pm0.1$     & 2.54 & 2.0\\
  \hline
$\tau^- \to e^+ \mu^- \mu^-$   & $9.2 \pm 1.1$ & 0 & $< 0.3$               & 2.55 & 2.0\\
  \hline
$\tau^- \to \mu^- e^- e^+$     & $9.4 \pm 0.8$ & 0 & 0.2$\pm0.2$     & 2.49 & 1.9\\
  \hline
$\tau^- \to \mu^+ e^- e^-$     & $9.5 \pm 1.4$ & 0 & $< 0.2$   & 2.55 & 2.0\\
  \hline
$\tau^- \to \mu^- \mu^+ \mu^-$ & $9.0 \pm 1.6$ & 0 & 0.1$\pm0.1$     & 2.51 & 2.0\\
  \hline
  \hline
 \end{tabular}\\
\\
\label{tab:result}
\end{table}

We determine the upper limit $s_0$ on the number of 
signal events at 90\% CL using the prescription of 
Feldman and Cousins~\cite{Poisson-limit}. To include 
in this limit the uncertainty in the detection efficiency 
$\epsilon$, we increase $s_0$ according to the prescription 
of Cousins and Highland \cite{Cousins-Highland}.
The main systematic uncertainties in the detection efficiencies come 
from tracking (1.0\% per track), electron identification (1.1\% per electron), 
muon identification (5.4\% per muon), trigger efficiency (1.4\%), 
statistics of signal MC (1.0\%) and uncertainty of the decay angular distribution 
($0.8-12.6$\%). The total uncertainty of the detection efficiency is 
6.1\% for  $\tau^- \to e^- e^+ e^-$,  15.1\% for  
$\tau^- \to e^- \mu^+ \mu^-$, 12.4\% for 
$\tau^- \to e^+ \mu^- \mu^-$, 8.4\% for 
$\tau^- \to \mu^- e^- e^+$, 15.1\% for 
$\tau^- \to \mu^+ e^- e^-$ and 17.5\% for 
$\tau^- \to \mu^-\mu^+\mu^-$ mode. The uncertainty in the 
number of $\tau$-pair events comes from the luminosity measurement
(1.4\%).  
For our calculation of $s_0$, we take the background (and its 
uncertainty) to be zero. This results in conservative upper limits.
Upper limits for branching fractions $\mathcal{B}$ are calculated 
for each decay mode as follows:
$\mathcal{B}(\tau^{-} \to \ell^{-}\ell^{+}\ell^{-}) < \displaystyle \frac{s_0}{2 N_{\tau\tau} \times \epsilon \times \mathcal{B}_1}$, 
where $N_{\tau\tau}$ is the total number of the $\tau$-pairs produced, 
and $\mathcal{B}_1$ is the 
inclusive 1-prong branching fraction of the $\tau$.
 The values of $s_{0}$ used and the resulting upper limits for the 
branching fractions are summarized in Table~\ref{tab:result}. 

\section{Summary}
We have searched for lepton-flavor-violating decays $\tau^- \to \ell^{-}\ell^{+}\ell^{-}$ 
using an 87.1 fb$^{-1}$ data sample. 
No evidence for any of these decay modes is observed 
and upper limits for the branching fractions are obtained in
the range $(1.9-3.5) \times 10^{-7}$ for 
$\tau^- \to \ell^{-}\ell^{+}\ell^{-}$ modes; 
these are approximately one order of magnitude more 
restrictive than the limits previously obtained by CLEO~\cite{CLEO3L}
and comparable to the recent results from BaBar~\cite{BABAR3L}. 

%***** Acknowledgments *****
\vspace*{1cm}
We wish to thank the KEKB accelerator group for the excellent
operation of the KEKB accelerator.
We acknowledge support from the Ministry of Education,
Culture, Sports, Science, and Technology of Japan
and the Japan Society for the Promotion of Science;
the Australian Research Council
and the Australian Department of Education, Science and Training;
the National Science Foundation of China under contract No.~10175071;
the Department of Science and Technology of India;
the BK21 program of the Ministry of Education of Korea
and the CHEP SRC program of the Korea Science and Engineering Foundation;
the Polish State Committee for Scientific Research
under contract No.~2P03B 01324;
the Ministry of Science and Technology of the Russian Federation;
the Ministry of Education, Science and Sport of the Republic of Slovenia;
the National Science Council and the Ministry of Education of Taiwan;
and the U.S.\ Department of Energy.

\end{document}